\documentclass[prd,aps, 11pt]{revtex4}
\usepackage{latexsym}
\usepackage{amssymb}
\usepackage[dvips]{graphicx}
\begin{document}
\newcommand{\beq}{\begin{equation}}
\newcommand{\eeq}{\end{equation}}
\newcommand{\ove}{\overline}
\newcommand{\half}{\frac 1 2 }
\newcommand{\fourth}{\frac 1 4}
\newcommand{\Fstar}{\raisebox{.2ex}{$\stackrel{*}{F}$}{}}
\newcommand{\Pstar}{\raisebox{.2ex}{$\stackrel{*}{P}$}{}}
\newcommand{\xv}{\mbox{\boldmath$\tau$}}
%
\newcommand{\et}{{\em et al}}
\newcommand{\ie}{{\em i.e.$\;$}}
\newcommand{\call}{{\cal L}}
%
%
\newcommand{\Prd}{Phys.  Rev. D$\;$}
\newcommand{\Prl}{Phys.  Rev.  Lett.}
\newcommand{\Plb}{Phys.  Lett.  B}
\newcommand{\Cqg}{Class.  Quantum Grav.}
\newcommand{\Np}{Nuc.  Phys.}
\newcommand{\Grg}{Gen.  Rel.  \& Grav.$\;$}
\newcommand{\Fp}{Fortschr.  Phys.}
\newcommand{\Sch}{Schwarszchild$\:$}
\renewcommand{\baselinestretch}{1.2}

\title{A classification of the effective metric 
in nonlinear electrodynamics}
\author{\'Erico Goulart de Oliveira Costa}
\affiliation{Centro Brasileiro de Pesquisas Fisicas,
Rua
Xavier Sigaud, 150, CEP 22290-180, Rio de Janeiro, Brazil.}
\author{Santiago Esteban Perez Bergliaffa}
\affiliation{
Departamento de F\'{\i}sica Te\'{o}rica,
Instituto de F\'{\i}sica,
Universidade do Estado de Rio de
Janeiro, CEP 20550-013, Rio de Janeiro, Brazil.}

\date{\today}
\vspace{.5cm}

\begin{abstract}
We show that only two 
types of effective metrics are possible in 
certain nonlinear electromagnetic theories.
This is achieved by using 
the dependence of the effective metric on the
energy-momentum tensor of the background along with
the
Segr\`e classification of the latter. 
Each of these forms is completely 
determined by 
single scalar function, which characterizes the 
light cone of the nonlinear theory. We compare this light cone 
with that of Minkowski
in two examples.

\end{abstract}

\maketitle



\vskip2pc

\section{Introduction}

The idea that gravity may be an emergent phenomenon 
\footnote{By emergence we understand
that in the systems in question \cite{sil} ''short-distance physics is radically different from long-distance physics''.}
described by an effective low-energy theory which a is consequence of averaging over (yet unknown) microphysical
degrees of freedom dates back at least to the proposal of Sakharov \cite{sakha} in 1968. Since then, it has been shown
that
is not difficult to get a low-energy effective metric (although it is considerably harder to get dynamical equations controlling it, see for instance \cite{matt0}). In fact,
the effective metric arises in systems of very different types (see \cite{mattlr} for a review). The basic idea behind this
generality has been exposed in \cite{matt0},
where it was shown that given
a classical single-field theory described by a Lagrangian that is an arbitrary  function of the field and its derivatives, the fluctuations of the linearized theory
around a non-trivial background propagate in a curved spacetime. The geometry of this
spacetime is encoded by
the effective metric, which is unique in the case of a single field, and depends on the background field configuration.
This feature of nonlinear theories led to the construction of
analog models of gravity, which imitate the kinematical properties of
gravitational fields (see \cite{mattlr} for a complete list).


In this article we shall show that the relation between the effective metric
in
nonlinear electromagnetic theories and the corresponding energy-momentum tensor
can be used   
to classify the possible metrics in terms of the Segr\`e types of $T_{\mu\nu}$ (see Sects.
\ref{rel}, \ref{algeb} and \ref{eig}).
We will see that for a given type of metric, the form of the light cone is 
univocally described by a scalar function, that depends on the Lagrangian of the theory and the background field
configuration (Sect.\ref{seco}).
Following the variation of the light cone along a given path, we can see how the particle is diverted
from the background geodesic motion due to the nonlinearities of the interaction.
Some examples of this application are given in Sect.\ref{simple}.
Finally we will discuss our results and consider future work in Sect.\ref{conc}.

\section{The effective geometry for nonlinear electromagnetism}
\label{rel}

Nonlinear electrodynamics is relevant in several areas of physics.
In quantum field theory, the polarization
of the vacuum leads naturally to a nonlinear corrections to Maxwell's electrodynamics,
which are described by Euler-Heisenberg's Lagrangian \cite{dunne}.
In material media, such as some
dielectrics and crystals, the complex interaction between the molecules and external electromagnetic fields can be described by an effective nonlinear theory, which is typically observed at very high light intensities such as those provided by pulsed lasers \cite{shen}. 

The result that
the high-energy
perturbations of a nonlinear electromagnetic theory propagate along
geodesics that are not null in the background geometry
but in an effective spacetime has been obtained several times in the literature \cite{nlemef}.
In the case of 
${\cal L}={\cal L}(F)$, where $F\equiv F_{\mu\nu}F^{\mu\nu}$ ,
the equation of motion is given by
\beq
({\cal L}_F F^{\mu\nu})_{,\nu}=0.
\eeq
By perturbing this equation around a fixed background solution and taking the eikonal limit
(see for instance \cite{matt1}),
we obtain for the effective metric
\beq
g^{\mu\nu} = {\cal L}_{F0} \eta^{\mu\nu}-4{\cal L}_{FF0}F^\mu_{\;\alpha 0} F^{\alpha\nu}_0,
\eeq
where the subindex 0 means that the quantity is evaluated using the background solution. 
Both this and the inverse metric can be expressed in terms of the energy-momentum tensor, given by
\begin{equation}
T^{\mu}_{\phantom a\nu}=-4{\cal L}_FF^{\mu}_{\phantom a\alpha}F^\alpha_{\phantom a\nu}-{\cal L} \delta^{\mu}_{\phantom a\nu}.
\end{equation}
in particular, the inverse effective metric takes the form
\begin{equation}
 g_{\mu\nu}=a_0\eta_{\mu\nu}+b_0T_{\mu\nu 0},
 \label{emem}
\end{equation}
where $a_0$ and $b_0$ are given by
\begin{equation}
a_0=-b_0\left[ \frac{{\cal L}_F^{2}}{{\cal L}_{FF}}+{\cal L}+\frac{1}{2}T\right]_0,
\label{azero}
\end{equation}
\begin{equation}
 b_0=\frac{16{\cal L}_{FF0}}{{\cal L}_{F0}}\left[ \kappa^{2}{\cal L}_{FF}^{2}-16({\cal L}_F+F{\cal L}_{FF})^{2}\right]^{-1}_0.
\label{bzero}
\end{equation}
with $\kappa = \sqrt{F^2+G^2}$, and
$G\equiv
F_{\mu \nu}^{*}F^{\mu\nu}$.
We shall discuss next how to use the dependence of the effective metric with $T_{\mu\nu}$ to classify the different possibilities
for $g_{\mu\nu}$.

\section{Algebraic properties of $T_{\mu\nu}$}
\label{algeb}
Due to the relation between the effective geometry and the energy-momentum tensor, the algebraic properties of
$T_{\mu\nu}$ determine the propagation of the high-energy perturbations of a given nonlinear theory. These properties
can be exhibited as a typical eigenvalue problem. Hence
it is useful to review some basic results concerning this problem in the context of relativity.

At a given point $p$ of a manifold $\mathcal{M}$, the object $T^{\alpha}_{\phantom a\beta}$ can be thought as a linear map of the tangent space $T_p$ onto itself. The principal directions of this map and its correspondent
eigenvalues are determined by
\footnote{To be precise, the eigenvalue problem for the energy momentum tensor
can be viewed as an immediate consequence of the
extremization of a scalar function
$\chi={T_{\mu\nu}\xi^\mu\xi^\nu}/{g_{\alpha\beta}\xi^\alpha\xi^\beta}$
with respect to $\xi^\alpha$ (in the same way as Petrov's classification of the
Weyl tensor is based on the extremization of the sectional curvature function).
In fact, deriving  $\chi$ with respect to $\xi^\alpha$ and imposing the condition of
extremum implies, the eigenvalue equation (5) results.}
\begin{equation}
T^{\alpha}_{\phantom a\beta}\xi^{\beta}=\lambda \xi^{\alpha},
\end{equation}
where $\lambda$ is a scalar and $\xi^{\beta}$ is an eigenvector. A fourth order characteristic polynomial for $\lambda$ is obtained by the condition $p(\lambda)={\rm det}(\lambda \textbf{1}-\textbf{T})=0$ \footnote{We sometimes use the symbols
$\textbf{T}$ and $\textbf{1}$
as matricial versions of the mixed tensors
$T^{\alpha}_{\:\beta}$ and
$\delta^{\alpha}_{\:\beta}$. Successive contractions of $T^{\alpha}_{\:\beta}$ will be denoted as
powers of $\textbf{T}$, i.e: $\textbf{T}^{2}\doteq T^{\mu}_{\:\alpha}T^{\alpha}_{\:\nu}$,
$\textbf{T}^{3}\doteq T^{\mu}_{\:\alpha}T^{\alpha}_{\:\beta}T^{\beta}_{\:\nu}$
and so on.}.
Although the algebraic properties of this equation are well known, it
is important to note that in a positive definite metric,
a real symmetric matrix can always be diagonalized by a real orthogonal transformation.
However, the hyperbolic character of a Lorentzian metric leads to a more complicated
algebraic situation. In particular,
the eigenvectors $\xi^{\beta}$ do not necessarily constitute a
linearly independent set, implying that $\textbf{T}$ may not have a
diagonal representation.
Notwithstanding this undesirable property, it is always possible to
reduce the matrix $\textbf{T}$ to a typical canonical form,
as will be briefly discussed in Sec.\ref{revis}.


\subsection{The Segr\`e classification revisited}
\label{revis}
The Segr\`e classification is a local, invariant and algebraic classification of
arbitrary second rank symmetric tensors (see for instance \cite{Hall2}).
Tensors of this type play a very
important role several areas of physics, and
the coordinate-independent method provided by Segr\`e has been
discussed by many authors in different contexts
\cite{Bona}.

It is possible to show that, depending on the properties of the characteristic polynomial $p(\lambda)$ 
given by 
\begin{equation}
p(\lambda) = \lambda^{4} -a_3\lambda^{3} +a_2\lambda^{2}-a_1\lambda+a_0=0,
\label{char}
\end{equation}
where the coefficients $a_n (n=0..3)$
are simple functions
of the scalar invariants built with powers of $\textbf{T}$,
and the nature of the eigenvectors, there exist at most four different classes (known as Segr\`e types) of symmetric tensors \cite{Hall2}. Those belonging to Segr\`e types I, II and III have only real eigenvalues and admit respectively, four, three or two linearly independent eigenvectors. Type IV describes tensors associated to complex and conjugated eigenvalues
\cite{Bona}. Each type is associated to a canonical form for the corresponding tensors.
By construction, type I is the only Segr\`e type in which the given linear map
admits a diagonal representation, and it can be shown that this is the only class that admits a timelike eigenvector
\cite{Bona, JSantos1}.
We list next two important properties that will be useful below \cite{Hall1}:\\[0.3cm]
\noindent (i) There always exists a two-dimensional subspace $S_p$ of $T_p$ which is invariant under the action of $\textbf{T}$\\
\noindent (ii) If $S_p$ is an invariant 2-space under the action of $\textbf{T}$, then so is the 2-space orthogonal to $S_p$.\\
The subspace \noindent$S_p$ will be called timelike, null or spacelike if it contains exactly two, one or no null vectors.

We shall see next how the Segr\`e classification can be useful in the determination of
possible types of effective metrics, in the particular example of nonlinear electromagnetism.

\section{Nonlinear electrodynamics and the eigenvalue problem of the energy-momentum tensor}
\label{eig}

\subsection{Linear case}

In order to study the properties of the energy-momentum tensor in nonlinear electromagnetism, we shall review first those of the linear theory, which furnishes a
simple realization of the Segr\`e classification \cite{Synge}. In this case, the energy-momentum tensor is given by
\begin{equation}
\tau^{\mu}_{\phantom a\nu}=F^{\mu}_{\phantom a\alpha}F^{\alpha}_{\phantom a\nu}+\frac{1}{4}F \delta^{\mu}_{\phantom a\nu}
\end{equation}
Beyond its obvious symmetric nature, $\xv$ satisfies
additional algebraic properties, such as 
tr$(\xv)=0$, and ${\rm tr}({\xv}^{2})  =  \tau^{\mu}_{\phantom a\nu}\tau^{\nu}_{\phantom a\mu}=\frac{1}{4}\kappa^{2}$,
where $\kappa\equiv (F^{2}+G^{2})^{1/2}$. 
It is easily shown that the characteristic equation (\ref{char}) for Maxwell's energy-momentum tensor becomes
\begin{equation}
\left(\lambda^{2}-\frac{\kappa^{2}}{16}\right)^{2}=0
\end{equation}
From this relation, two important results follow:\\
(i) Since the characteristic equation factors in two identical second order polynomials, the four eigenvalues of the electromagnetic energy-momentum tensor in linear electromagnetism are real and equal in pairs.\\
(ii) The eigenvalues at each point of spacetime are entirely described by a given function of the field invariants $F$ and $G$ at that point:
\begin{equation}
\lambda_\pm=\pm\frac{1}{4}\kappa
\end{equation}
We shall consider next the eigenvector structure,
which can be determined by the analysis of the only two possible cases for $\kappa$ ($\kappa\neq 0$ and $\kappa =0$) \cite{Synge, Kramer}. 
Note that because there exist at most two eigenvalues, the structure of a given %
eigenspace is degenerate, and there is an infinite number of $\xi^{\alpha}$ associated to a given $\lambda$.
\subsubsection{Non-null field ($\kappa\neq0)$: two different eigenvalues.}
In this case, there exist two orthogonal invariant eigenspaces $S_p$, each of them in correspondence with a given eigenvalue $\lambda_{\pm}$. It also follows that the timelike two-flat (which admits exactly two null eigenvectors) is associated to the positive eigenvalue $\lambda_{+}$ \cite{Synge}. Because the other two-flat admits two orthogonal spacelike vectors, it is possible to diagonalize $\xv$ and it follows that non-null Maxwell fields are of Segr\`e type I.

\subsubsection{Null field ($\kappa =0$): one single null eigenvalue.}
It is possible to show that in this case the eigenvectors of $\xv$ do not constitute a linearly independent set. Nevertheless, there exist a three-flat in $T_p$ that admits a null direction and two independent spacelike vectors. This three-flat is tangent to the light cone at $p$. According to the Segr\`e classification, this case belongs to type II. Hence $\xv$ does not admit a diagonal representation but can be reduced to a
canonical form (which will be given in Sec.\ref{st2}) \cite{corm}.

\subsection{Nonlinear electromagnetism}

We now turn to the algebraic analysis of the energy-momentum tensor constructed with nonlinear and real Lagrangians for the electromagnetic field. As will be shown below, such analysis will be very useful in the description of the light cone structure
of an arbitrary nonlinear electromagnetic theory.
First we will consider 
Lagrangians that are
arbitrary functions of the electromagnetic invariants $F$ and $G$.
Following the standard defnition of the energy-momentum tensor \cite{Landau}, we get
\begin{equation}
T^{\mu}_{\phantom a\nu}=-4{\cal L}_FF^{\mu}_{\phantom a\alpha}F^\alpha_{\phantom a\nu}-({\cal L}-{G\cal L}_G)\delta^{\mu}_{\phantom a\nu},
\end{equation}
where ${\cal L}_A\equiv \partial {\cal L}/\partial A$, $A=F,G$.
The roots of the characteristic
polynomial $p(\lambda)$ (Eqn.(\ref{char}))
would give the spectrum
of
eigenvalues for a given configuration of fields in the context of a given nonlinear
theory.
However, a simple inspection  of the symmetries of $\textbf{T}$
reveals a much more complicated structure than that of Maxwell's.
First of all, the trace of $T^{\mu}_{\phantom a\nu}$ does not vanish
and is given by tr$(\textbf{T})=-4({\cal L}-F{\cal L}_F-G{\cal L}_G)$
Furthermore,
a calculation of the powers of \textbf{T} reveals that, for a given integer $n$
\begin{equation}
\textbf{T}^n=\alpha\textbf{F}^{2}+\beta\textbf{1},
\end{equation}
with $\alpha$ and $\beta$ are functions of the
invariant $F$ and $G$, and of ${\cal L}$ and its derivatives. Thus, the nonlinearity of the theory deforms the Rainich algebra \cite{Rainich} $\textbf{\xv}^{2}\sim\textbf{1}$ valid in the linear theory, making the calculation of the coefficients of the characteristic polynomial 
hard.
Nevertheless, it is possible to bypass this difficulty by
decomposing $\textbf{T}$ in terms of its traceless part $\textbf{N}$ and its trace, that is
$$
N^{\alpha}_{\;\;\beta}=T^{\alpha}_{\;\; \beta}-\frac{1}{4}T\delta^{\alpha}_{\;\; \beta}.
$$
Then, if $\mathbf{\xi}$ is an eigenvector of $\textbf T$ with eigenvalue $\lambda$ it will be also an eigenvector of $\textbf N$ with eigenvalue $\lambda-T/4$. Hence we are led to the study the properties of $N^{\alpha}_{\phantom a\beta}$ which for an
arbitrary Lagrangian is given by
\begin{equation}
N^{\alpha}_{\phantom a\beta}=-4{\cal L}_F\left(
F^{\alpha}_{\phantom a\lambda}F^{\lambda}_{\phantom a\beta}+\frac{1}{4}
F\delta^{\alpha}_{\phantom a\beta}\right).
\end{equation}
In other words, the traceless part of the nonlinear energy-momentum tensor is just a conformal transformation of Maxwell's $\xv$, and the eigenvalues of $\textbf{N}$ are just that of $\xv$
multiplied by $-4{\cal L}_F$
This interesting fact permits to obtain the eigenvalues of \textbf{T} with the expression
\begin{equation}
\lambda_{\pm}=F{\cal L}_F+G{\cal L}_G-{\cal L}\mp {\cal L}_F \kappa.
\end{equation}
Thus, the eigenvalues are given in terms of the field invariants and specific functions of them. 
Notice that 
the first three terms are just the trace of the energy-momentum tensor divided by four. Furthermore, the invariant subspaces of Maxwell's theory determine entirely the invariant subspaces of a nonlinear theory. We conclude that even in a nonlinear theory with $\call = \call (F,G)$, the algebraic properties of the energy-momentum tensor are such that the only possible Segr\`e types are I and II. Furthermore, type II is only possible if $\kappa =0$. In the next section we show how the Segr\`e types determine the light cone structure in
the particular example of nonlinear electromagnetism.

\section{Second order surfaces and a classification of nonlinear regimes}
\label{seco}
The light cone structure of a nonlinear theory is governed by the effective metric
$g_{\mu\nu}$ through the condition
\begin{equation}
g_{\mu\nu}k^\mu k^\nu=0,
 \label{lc}
\end{equation}
where the $k^\mu$ are null vectors in the effective geometry but are not null, in general, in the background geometry (which maybe flat or curved). We shall study next the the relation between the newly-defined light cones and the background light cones (which for definiteness
we assume to be those of Minkowski spacetime). It will be shown that there exist many possibilities that are associated with the algebraic nature of the energy-momentum tensor at a given spacetime point, and to the properties of the Lagrangian. Notice that although $g_{\mu\nu}$ 
can be taken as a new Riemannian metric, it is also licit to think of it
as a tensor field defined in Minkowski spacetime. Because of the expression
$$
g_{\mu\nu}=a_0\gamma_{\mu\nu}+b_0T_{\mu\nu 0},
$$
with $a_0$ and $b_0$ given by Eqns.(\ref{azero}) and (\ref{bzero}),
the principal directions (eigenvectors) of $g_{\mu\nu}$  and its invariant subspaces are entirely determined by the eigenvectors of $T^{\mu}_{\phantom a\nu}$.  The analysis will be divided in two different classes, following the two different algebraic types of the energy-momentum tensor discussed above.

\subsubsection*{Class 1: Segr\`e type I background}

If, at a given point $p$, the energy-momentum is of Segr\`e type I, the following lemma ensues:\\[0.3cm]
\noindent \textit{Lemma:} There exist a non-null intersection between Maxwell's light cone and the nonlinear light cone, which is given by the null eigenvectors of $T^{\mu}_{\phantom a\nu}$.

\noindent Proof: As stated before, in Segr\`e type I there are two eigenvectors of $\mathbf T$ (those with eigenvalue $\lambda_+$) that are null in the background metric. Denoting them by $\xi^\mu_{(i)}$, $i=1,2$, we have
$$
g_{\mu\nu}\xi^\mu_{(i)}\xi^\nu_{(i)}=a_0\gamma_{\mu\nu} \xi^\mu_{(i)}\xi^\nu_{(i)}+
b_0\xi^\mu_{(i)}T_{\mu\nu}\xi^\nu_{(i)}= (a_0+\lambda_+b_0)\gamma_{\mu\nu} \xi^\mu_{(i)}\xi^\nu_{(i)} =0.$$
Since we have $g_{\mu\nu}\xi_{(i)}^{\mu}\xi_{(i)}^{\nu}=0$ and $\eta_{\mu\nu}\xi_{(i)}^{\mu}\xi_{(i)}^{\nu}=0$, the
$\xi_{(i)}^{\mu}$ are
contained in both light cones, proving that they are tangent along these vectors.

By choosing a specific orthogonal basis at $p$, the energy-momentum tensor admits a diagonal representation. which allows the light cone condition
to be written as
\begin{equation}
f_{+}[(k^0)^{2}-(k^1)^{2}]-f_{-}[(k^2)^{2}+(k^3)^{2}]=0
\label{cono1}
\end{equation}
where $f_\pm=a_0+b_0\lambda_\pm$ are the eigenvalues of the effective metric tensor given in Eqn.(\ref{emem}).
At a given point $p$, this is the equation of a three dimensional surface representing locally the light cone of a nonlinear theory in the tangent space $T_p$. The coefficients $f_{+}$ and $f_{-}$ depend explicitly on the Lagrangian and its derivatives, and on the background field,
but in a sense, for a given theory, the three-surface
will depend only on the invariant $F$. In the context of Maxwell's linear theory this equation reduces to that of the light cones of Minkowski spacetime.

\begin{figure}[h]
\begin{center}
\includegraphics[width=0.6\textwidth]{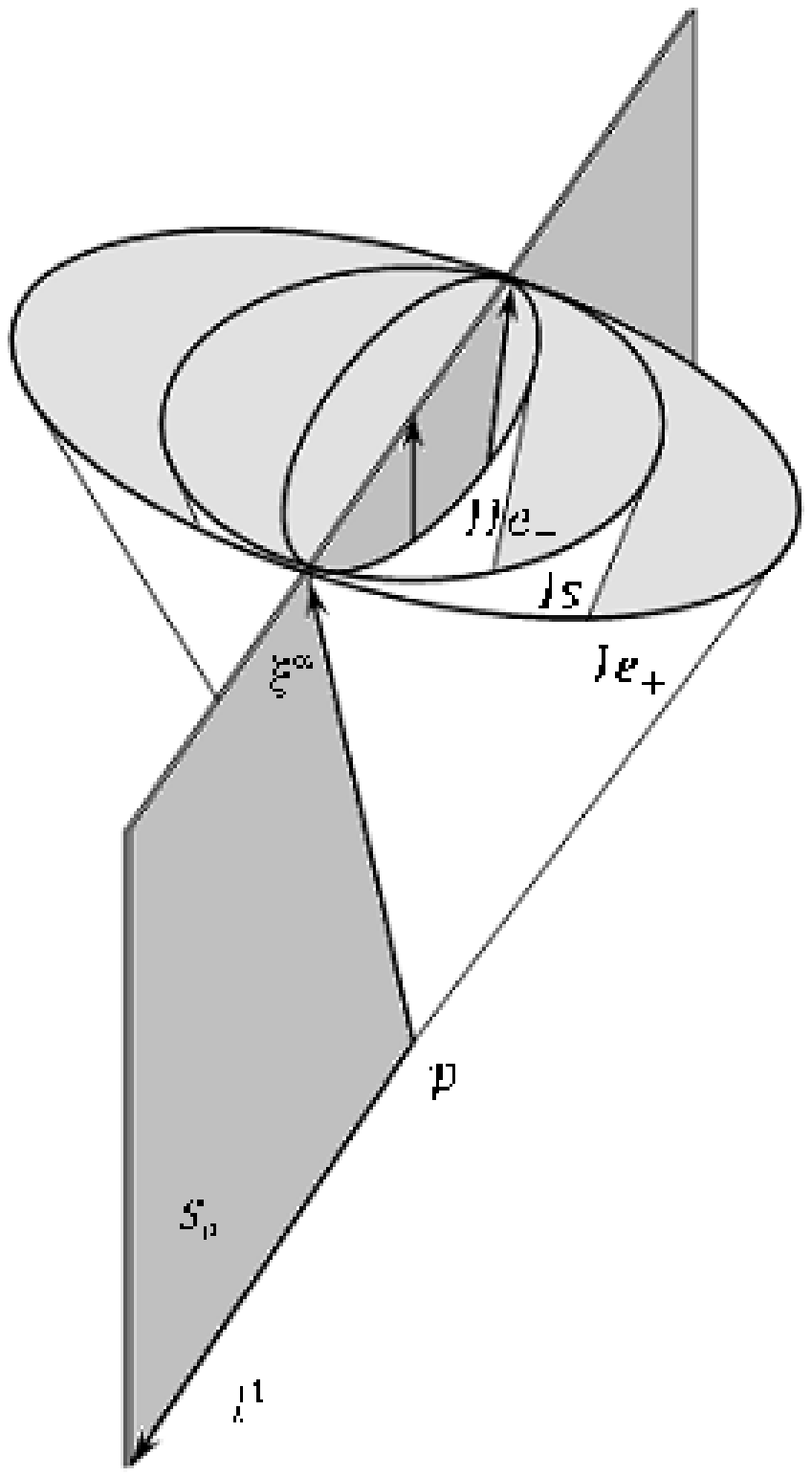}
\caption{Minkowski and nonlinear light cones for Segr\`e type I and $\Upsilon >0$.}
\label{s1}
\end{center}
\end{figure}

To develop a geometrical understanding of the causal structure, we will set $k^0=1$ in Eqn.(\ref{cono1}). From a physical point of view, we can interpret the resulting bidimensional surface as the position of the wavefront
after a given infinitesimal lapse of time. In the case of linear electromagnetism, this surface is a sphere (called
Maxwell's sphere from now on). Because there exist two common null vectors to both surfaces, the two-dimensional surface that follows from Eqn.(\ref{cono1})(which we will call
the photon surface)
intersects Maxwell's sphere precisely at two points ($k_{1}=\pm1$ in this frame).
Defining the function $\Upsilon\equiv f_{-}/f_{+}$ we have
\begin{equation}
(k^1)^{2}+\Upsilon [(k^2)^{2}+(k^3)^{2}]=1.
\end{equation}
From the theory of quadrics we have the following possibilities for the resulting photon surface. If $\Upsilon>0$,
the surface is an ellipsoid of revolution around the $k^1$ axis. The ellipsoid touches Maxwell's sphere from inside
if $\Upsilon>1$ (with the major axis along $k^1$) and from outside
if $0<\Upsilon<1$ (with minor axis along $k^1$, see Fig.\ref{s1}). The difference between the two light cones shows that
because of the interaction with the
background field, the nonlinear photons (NLP)
propagate with different velocities in different directions. In the first case,
all the NLP (except those along $k^{1}$) propagate with velocities that are less than $c$, while in the second
case all the NLP (leaving aside those along $k^{1}$) propagate with velocity greater that $c$. The limiting case $\Upsilon=1$ coincides with Maxwell's sphere, in such a way that any theory with a background such that $\Upsilon =1$ at a
given point will locally reproduce every property of Maxwell's theory from the point of view of photon propagation.

\begin{figure}[h]
\begin{center}
\includegraphics[width=0.5\textwidth]{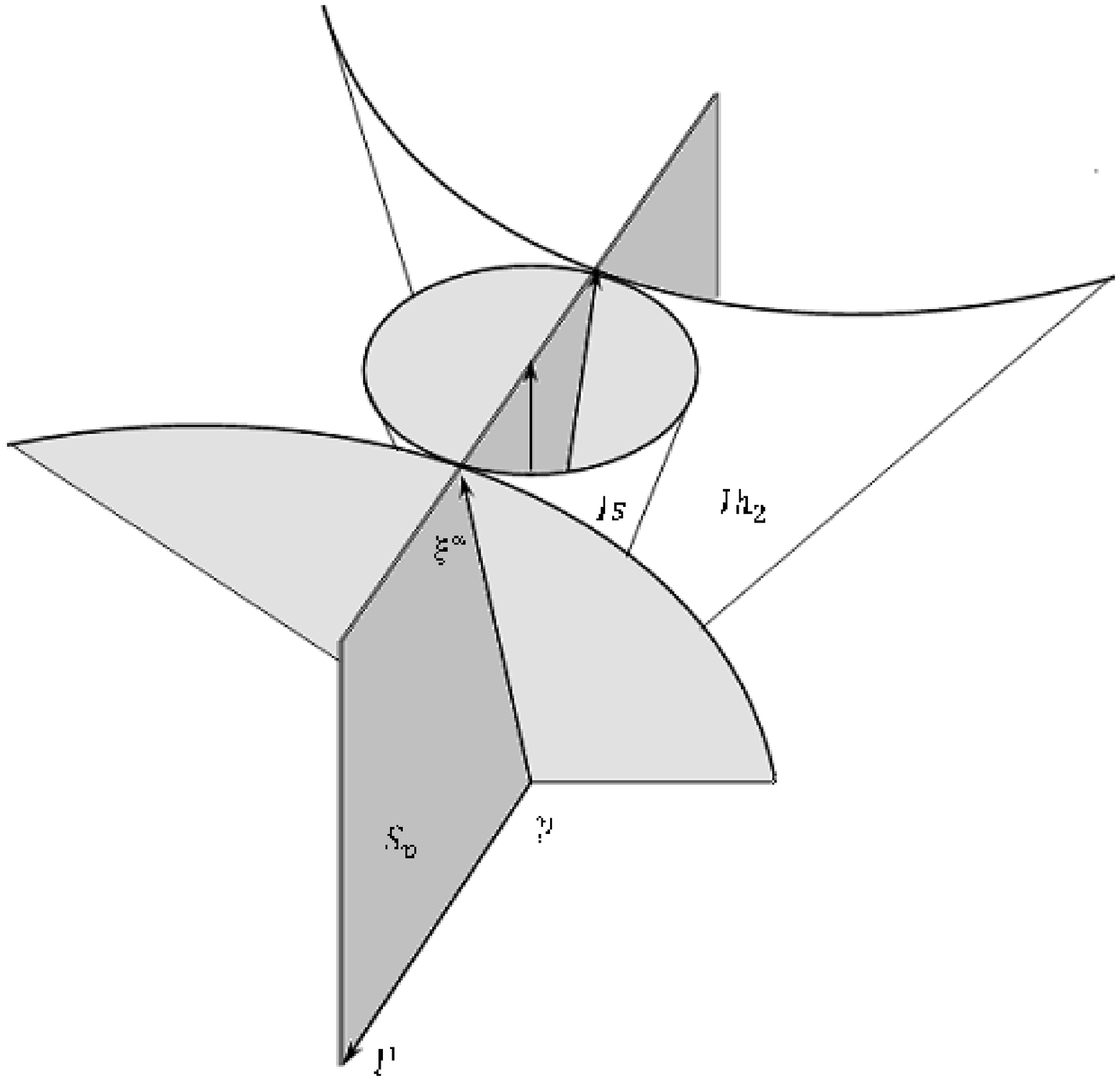}
\caption{Minkowski and nonlinear light cones for Segr\`e type I and $\Upsilon < 0$.}
\label{s1h2}
\end{center}
\end{figure}

\begin{figure}[h]
\begin{center}
\includegraphics[width=0.5\textwidth]{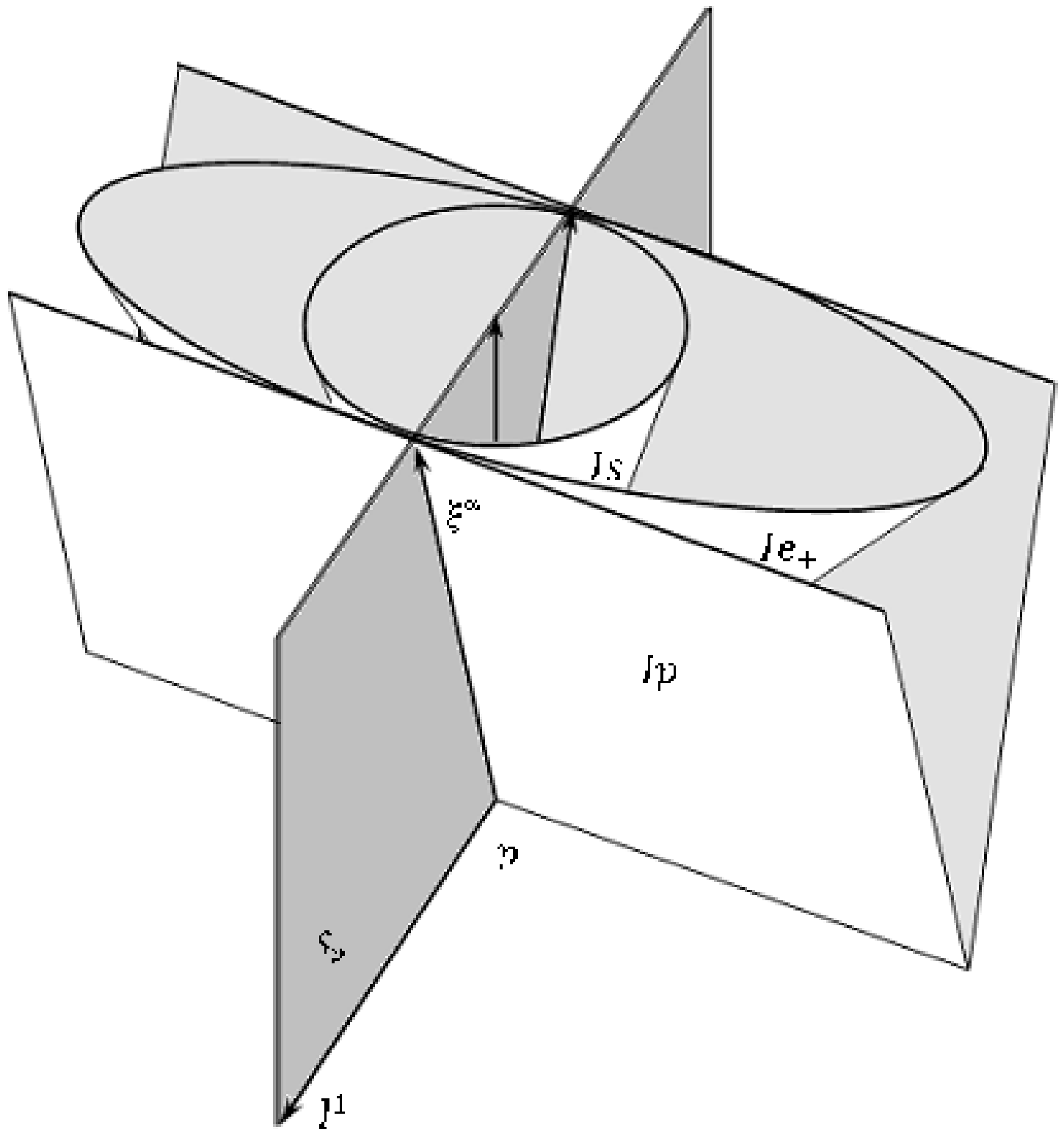}
\caption{Minkowski and nonlinear light cones for Segr\`e type I and $\Upsilon \geq 0$.}
\label{s1p}
\end{center}
\end{figure}

Notice that there exist two more exotic allowed situations. When the condition $\Upsilon=0$ is satisfied,
the photon surface is determined by planes satisfying the condition $k^1=\pm1$.
These planes intersect Maxwell's sphere in two points, in which case the velocity of the NLP coincides with $c$.
This in this case the nonlinear theory and the background are such that light propagation is
forbidden in spacelike directions which in the chosen frame, are orthogonal to the $k^{1}$ axis.
A second exotic possibility is
$\Upsilon<0$, since the three-surface will be a two-sheet hyperboloid touching the sphere from outside.
In this interesting case the spacelike directions in which the NLP cannot propagate are many more than in
the $\Upsilon=0$ case.
The following table displays the catalogue of possible light cone
structures for a Segr\`e I type background energy-momentum
tensor, along with a notation based on
the algebraic type and resulting photon surface.
\begin{table}[ht!]
 \centering
 \begin{tabular}{ccc} \hline\hline
$\Upsilon$ & $\mathrm{photon}\: \mathrm{surface}$ & ${\rm symbol}$ \\
\hline
$\Upsilon>1$ & ${\rm internal}\: {\rm ellipsoid}$ &$ Ie_{-}$ \\
$\Upsilon=1$ & ${\rm sphere}$ & $× Is$\\
$0<\Upsilon<1$ & ${\rm external}\:{\rm ellipsoid}$ & $Ie_{+}$  \\
$\Upsilon=0$ & ${\rm planes}$ & $×Ip$ \\
$\Upsilon<0$ & ${\rm two-sheet}\:{\rm hyperboloid}$ & $Ih_{2}$\\
\hline
 \end{tabular}
\caption{The table shows the different photon surfaces for a Segr\`e type I energy-momentum tensor.}
\label{t1}
\end{table}
A straightforward calculation shows that for $\call = \call (F)$,
\begin{equation}
\Upsilon=\frac{{\cal L}_F+{\cal L}_{FF}(F-\kappa)}{{\cal L}_F+{\cal L}_{FF}(F+\kappa)}.
\label{upsi}
\end{equation}
Thus, the local causal structure in any nonlinear theory of electrodynamics is determined by computing a single function of the field invariant.
The figures \ref{s1},\ref{s1h2}, and \ref{s1p} show some properties of the resulting surfaces and their relation with the Minkowski light cone.

\subsubsection*{Class 2: Segr\`e type II background}
\label{st2}
\begin{figure}[h]
\begin{center}
\includegraphics[width=0.65\textwidth]{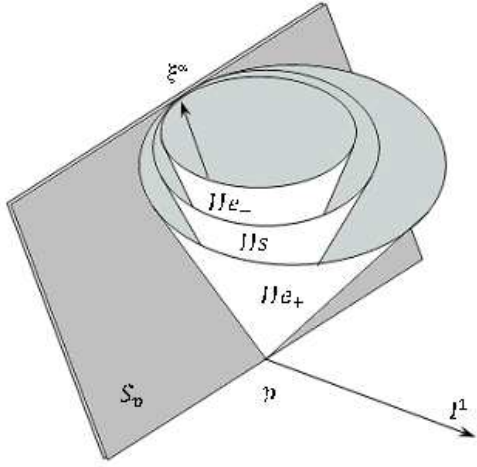}
\caption{Minkowski and nonlinear light cones for a Segr\'e type II energy-momentum tensor.}
\label{s2}
\end{center}
\end{figure}

In this case the energy-momentum tensor cannot be diagonalized and, because $F=0$, there exists only one eigenvalue $\lambda=-{\cal L}$. Nevertheless, it is always possible to reduce $\textbf{T}$
to the following matrix representation by choosing a specific orthogonal frame:
\begin{equation}
T^{\mu}_{\phantom a\nu}=\left(
\begin{array}{llll}
\lambda-\mu & -\mu & 0 & 0 \\
\mu & \lambda+\mu & 0 & 0 \\
0 & 0 & \lambda & 0 \\
0 & 0 & 0 & \lambda
                        \end{array}\right)
                        \end{equation}
where $\mu\equiv 2{\cal L}_F(E^2+H^2)$. Following the procedure sketched in the last subsection, \textit{i.e.} imposing the condition
$k^0=1$ and introducing $f_0\equiv a_0+b_0\lambda$, Eqn.(\ref{lc}) becomes
\begin{equation}
(\Phi+1)(k^1)^2+(k^2)^2+(k^3)^2+2\Phi k^1+(\Phi-1)=0
\end{equation}
where $\Phi\equiv b_0\mu_0/f_0$. Since there exists only one null eigenvector for a Segr\`e type
II energy-momentum tensor, it is immediate to show that the corresponding photon surface intersects
Maxwell's sphere only at one point ($k^1=-1$ in the chosen frame). Also, the non-diagonal terms in $\textbf{T}$
imply that the resulting surface is not centered at the origin. As in Segr\`e type I, we have several possibilities.
If $-1<\Phi<\infty$, the three-surface will be an ellipsoid of revolution along $k^{1}$.
When $0<\Phi<1$, the surface is entirely contained inside Maxwell's sphere, while in the case $-1<\Phi<0$
Maxwell's sphere is inside, in such a way that the limiting case $\Phi=0$ is  the sphere itself (see Fig.\ref{s2}).
There are also some exotic situations. If $\Phi=-1$, the NLP define a paraboloid of revolution around $k^{1}$, which prohibits propagation in some directions. Finally, the condition $\Phi<-1$ determines a strange situation in which a one-sheet hyperboloid
is tangent to the sphere. The different possibilities are given in the following table, where
for $\call = \call (F)$, $\Phi$ is given by
\begin{equation}
\Phi=-\mu \left.\frac{{\cal L}_{FF}}{{\cal L}_{F}}\right|_0.
\end{equation}
\begin{table}[ht!]
 \centering
 \begin{tabular}{ccc} \hline\hline
 $\Phi$ & $\mathrm{photon}\:\mathrm{surface}$ & $\mathrm{symbol} $\\
\hline
$0<\Phi<\infty$ & $\mathrm{internal}\;\mathrm{ellipsoid}$ & $IIe_{-} $\\
$\Phi=0$ & $\mathrm{sphere}$ & $IIs$ \\
$-1<\Phi<0$ &$ \mathrm{external}\:\mathrm{ellipsoid}$ &$ IIe_{+} $\\
$\Phi=-1$ & $\mathrm{paraboloid}$ & $IIp$ \\
$\Phi<-1 $& $\mathrm{one-sheet}\:\mathrm{hyperboloid}$ & $IIh_{1}$\\
\hline
\end{tabular}
\caption{The table shows the different photon surfaces for a Segr\`e type I energy-momentum tensor.}
\end{table}

\section{Simple realizations of the classification}
\label{simple}
In this section we will discuss the relation between the orientation of the 
light cones in a nonlinear $\call (F)$ theory
and a given electromagnetic field. For simplicity, we will concentrate only in Segr\'e I type energy-momentum tensors.
In terms of the electromagnetic vectors $E^{\:\mu}$, $H^\mu$ and a given timelike congruence
$V^{\mu}$, 
the electromagnetic field tensor admits the following
decomposition: 
$$
F^{\mu \nu}\equiv E^{\left[\mu\right.}V^{\left.\nu\right]}+\frac{1}{2}\eta^{\mu \nu}_{\phantom a \phantom a \alpha \beta}H^{\alpha}V^{\beta}
$$
where the electric and magnetic vectors $E^\mu(x)$ and $H^\mu(x)$
are defined respectively as $E^\mu = F^{\mu}_{\phantom a \nu}V^{\nu}$ and $H^{\mu}\equiv \eta^{\mu}_{\phantom a \varepsilon \alpha \beta}V^{\varepsilon}F^{\alpha \beta}$, and $F=2(H^{2}-E^{2})$, $G=-4E\cdot H$. The energy-momentum tensor for a theory described by $\call = \call (F)$ can be similarly decomposed as
\begin{eqnarray}
T_{\mu\nu} & = & -4\call_F(E^{2}V_{\mu}V_{\nu}-E_{\mu}E_{\nu}+2q_{(\mu}V_{\nu)}))\\
 & &-4\call_F(H^{2}V_{\mu}V_{\nu}-H_{\mu}H_{\nu}-H^{2}g_{\mu\nu})-{\cal L}g_{\mu\nu},
\end{eqnarray}
where $E^{2}=-E_{\mu}E^{\mu}$, $H^{2}=-H_{\mu}H^{\mu}$ and $q_{\lambda}=\frac{1}{2}\eta_{\lambda}^{\phantom a\mu\rho\sigma}E_{\mu}V_{\rho}H_{\sigma}$ is
the Poynting vector.
In Segr\'e type I tensors, it is always possible to find an observer such that the electric and
magnetic fields are parallel, \textit{i.e.} $H_{\mu}=\xi E_{\mu}$ and, consequently, the Poynting vector
vanishes \cite{Landau}. It is also possible to show that $V^{\mu}$ and $E^{\mu}$ are eigenvectors of
$T^{\alpha}_{\phantom a\beta}$ with eigenvalue $\lambda_{+}$. Thus, the timelike invariant subspace $S_p$
is spanned by a linear combination of $V^{\mu}$ and $E^{\mu}$. The spacelike subspace is determined by means of
two spacelike orthonormal vectors $n^{\mu}$ and $m^{\mu}$ satisfying $n^{\mu}E_{\mu}=m^{\mu}E_{\mu}=0$.
It is immediate to see that these vectors are eigenvectors of the energy-momentum tensor with eigenvalue $\lambda_{-}$. In the frame where $H_{\mu}=\xi E_{\mu}$,
the eigenvalues have the form:
\begin{eqnarray}
&&\lambda_{+}=-4{\cal L}_{F}E^{2}-{\cal L}\\
&&\lambda_{-}=4 {\cal L}_{F}\xi^{2}E^{2}-{\cal L},
\end{eqnarray}
In particular, if the observer is such that the magnetic part $H^{\mu}$
vanishes ($\xi =0$), the electric field direction determines completely the orientation of the timelike subspace which, as proved by the
lemma in Sec.\ref{seco},
determines the only directions in space in which light propagation is not affected by the nonlinear interaction. We will analyze next the problem of a static electrically charged particle in the context of two different nonlinear theories of electrodynamics, which illustrates what has been exposed here.

\subsection{Example 1}
Following the developments of the
previous section, we shall examine here
the light cone structure generated by the field of an electric charge placed at the origin in
Born-Infeld theory, with Lagrangian given by
$$
{\cal L}=b^2\left(1-\sqrt{1+\frac{F}{2b^2}}\right).
$$
The equations for the EM field,
$$
(\sqrt{-g}{\cal L}_FF^{\mu\nu})_{;\nu}=0
$$
in the spherically symmetric case have been solved for instance in \cite{wh}.
The only nonzero component of $F^{\mu\nu}$ is $F^{tr}$ in such a way that
$$
\frac{F}{2b^2}=-\frac{16\alpha^2}{r^4b^2+16\alpha^2},
$$
where $\alpha$ is proportional to the electric charge.
Defining the new variable
$$
y=\left(\frac{b^2r^4}{16\alpha^2}\right)^{1/4},
$$
the function $\Upsilon$ introduced in Eqn.(\ref{upsi}) is in this case given by
$$
\Upsilon = 1 - \frac{1}{1+y^4}.
$$
$\Upsilon$ tends to 1 for large $r$ (meaning that the linear theory is recovered far from the charge),
and it is actually restricted to the interval $[0,1]$.
From the classification in Table \ref{t1}, we see that the light cones
go from a sphere (for very large values of $r$) to an external ellipsoid, deforming
finally to two parallel planes (for $r=0$).
Note that these considerations are valid in the reference frame in which $H_\mu $ is zero.

This case has been discussed before in \cite{wh}, from the point of view of the effective potential for the NLP
arising from the effective metric, given by
$$
V(r) = \frac{L^2r^2}{r^4+\frac{8\alpha^2}{b^2}}.
$$
We see that for NLP travelling in the radial direction, $V=0$. Hence, these photons move
with velocity $c$, as was shown in Sec.\ref{seco}. The motion for $L\neq 0$ can be
described
in the formalism presented here by the analysis of the variation of the local light cone,
exactly as is done for instance in the case of the Schwarzschild's black hole.
Notice that all the necessary information about the light cones
is encoded in $\Upsilon$.

\subsection{Euler-Heisenberg}
The second example that will be examined here is that of an Euler-Heisenberg-like Lagrangian (in the weak-field
limit), given by
$$
\call (F) = -\frac 1 4 F+\beta F^2,
$$
where for the time being we leave $\beta$ unspecified.
Taking into account that $\kappa=+\sqrt{F^2+G^2}$,
we can distinguish two cases:

1) $F>0\rightarrow \kappa = F$. In this case,
$$
\Upsilon = \frac{8\beta F-1}{24\beta F-1}.
$$

\begin{figure}[h]
\begin{center}
\includegraphics[width=0.6\textwidth]{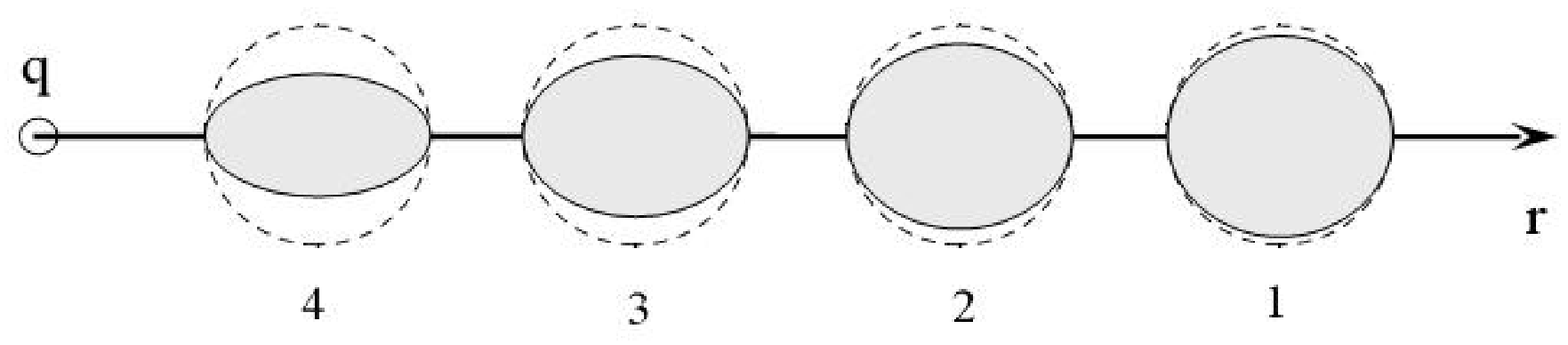}
\caption{Variation of Minkowski and photon spheres for $\beta >0$.}
\label{erico11}
\end{center}
\end{figure}

\begin{figure}[h]
\begin{center}
\includegraphics[width=0.6\textwidth]{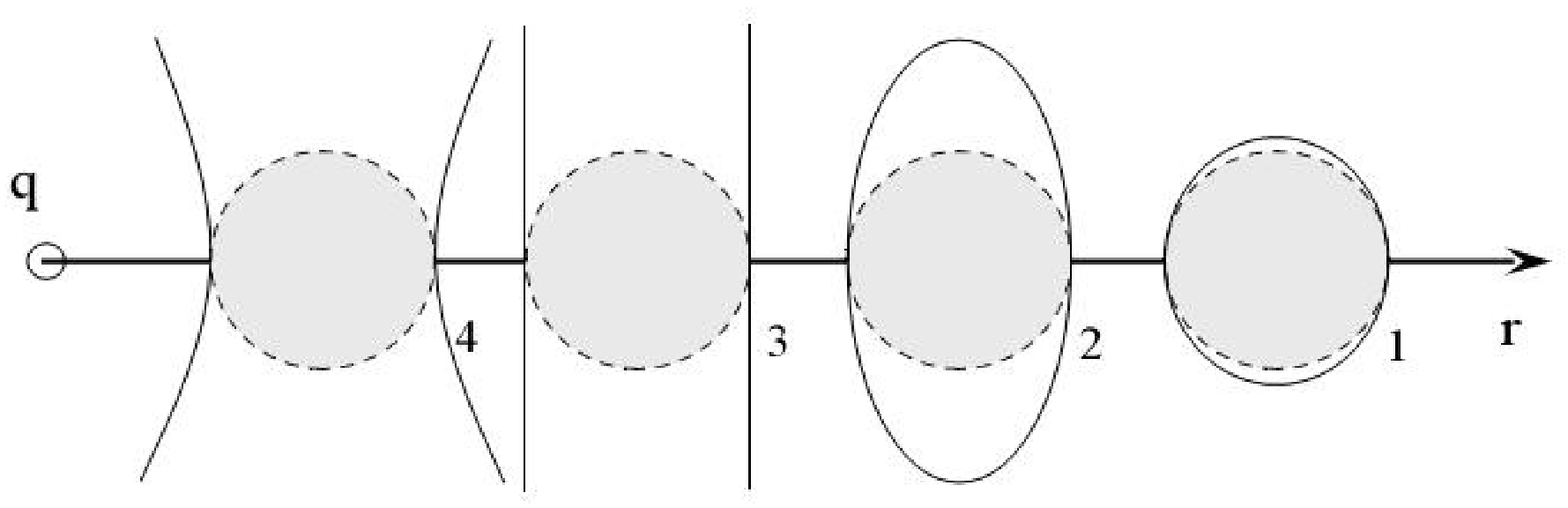}
\caption{Variation of Minkowski and photon spheres for $\beta >0$.}
\label{erico22}
\end{center}
\end{figure}

2) $F<0$, then $\kappa = - F$, and
$$
\Upsilon = \frac{24\beta F-1}{8\beta F-1}.
$$
On the other hand, the equations of motion
$$
(\call_FF^{\mu\nu})_{;\nu}=0,
$$
in the case of the EH-like Lagrangian for an electric charge
lead to
$$
\frac{E(r)}{4}+4\beta E(r)^3 = \frac{Q}{r^2},
$$
where $Q$ is proportional to the electric charge.
This expression can be inverted to give
$$
r^2=\frac{Q}{\sqrt{\frac E 4 + 4 \beta E^3}}.
$$
We see that for $\beta >0$, the electric field must
satisfy $0<E^2<\infty$.
It follows that in this case
$-\infty < F < 0$. From the plot of $\Upsilon$ in terms of $F$
it follows that $\Upsilon$ goes from the value 1 (at $r\rightarrow\infty$)
to the value 3 (at $r\rightarrow 0$). Hence the photon spheres
in this case vary qualitatively shown in Fig.\ref{erico11}.
In the same way,
for $\beta <0$, we have that $\Upsilon \rightarrow 1$ for $r\rightarrow\infty$, and
$\Upsilon \rightarrow -\infty$ for $\rightarrow 0$. The qualitative variation
of the photon spheres in this case is shown in Fig.\ref{erico22}
%
%
\section{Conclusion}
\label{conc}

We have shown that the presence of $\mathbf{T}$ in the effective metric along with the help of the Segr\`e classification of second rank symmetric tensors
furnishes a classification of the possible forms of the effective metric for nonlinear electromagnetic theories.
We developed this classification in the case of 
Lagrangians given by 
$\call = \call (F)$ \footnote{The generalization to $\call (F,G)$ Lagrangians should be straightforward, taking into 
account that the effective metric can still be expressed as in Eqn.(\ref{effmet}), see \cite{klo}}, 
showing that there are only two possible general forms for the effective metric (associated to Segr\`e
types I and II, which are the only types that appear in this case). The explicit form of the effective metric
can be used to compare the effective light cone with the Minkowskian light cone.
We have presented the different possibilities, each of them associated to 
a single scalar function of the Lagrangian,
its derivatives, and the background field. Finally, we have analyzed two examples,
which illustrate the power of the method, in the sense that
the variation of the light cones  is encoded in $\Upsilon$.

To close, let us remark that although we have focused in the case of the nonlinear electromagnetic field, the classification of the effective metric in terms of the Segr\`e types 
is possible for other fields. This follows from the fact that,
as shown in \cite{mariofierz} and \cite{mm}, the effective geometries 
associated to nonlinear scalar field and the nonlinear spin two field theories 
can also be written in the form 
\beq
g_{\mu\nu}=\Omega^{(1)}_0\gamma_{\mu\nu} +\Omega^{(2)} _0 T_{\mu\nu 0}.
\label{effmet}
\eeq
In this expression,  $\gamma_{\mu\nu}$ is the background metric, and 
$\Omega^{(1)}_0$ and $\Omega^{(2)}_0$ are functions of the background field, the detailed form of which depends on
the given theory, and $\Omega^{(2)}_0$ is such that it goes to zero when the theory
is linear. We also leave for future work half-integer spin fields, and the case of multi-fields.

\section{Acknowledgements}
SEPB would like to acknowledge support from ICRANet (where part of this 
work was done), and FAPERJ.

\end{document}